\documentclass{mem}
\usepackage{natbib}\usepackage{txfonts}\usepackage{balance}
\usepackage{graphicx}
\usepackage{txfonts}
\usepackage[a4paper]{hyperref}
\idline{79}{1}
\begin{document}
\def\teff{$T\rm_{eff }$}
\def\kms{$\mathrm {km s}^{-1}$}

\title{
CIAO? ADiOS!
}

   \subtitle{}

\author{
Andreas Korn
    }

\institute{
Department of Astronomy and Space Physics, Uppsala University, Box 515, 75120 Uppsala, Sweden;
\email{akorn@astro.uu.se}
}

\authorrunning{Korn }

\titlerunning{Atomic Diffusion in Old Stars}

\abstract{At the {\sl New Horizons in Globular Cluster Astronomy} conference (Padova, June 2002), two members of the {\sl VLT globular cluster team\/} presented different views on the importance of heavy-element sedimentation in Population II stars: ``The lack of evidence for depletion of Fe and Li in the atmospheres of globular cluster subgiants led {\em some people} to suspect that, for unknown reasons, Population II stars are not affected by this mechanism.'' \citep{Castellani_2003} and ``There should be {\em some mechanism} that prevents sedimentation.'' \citep{Gratton_2003}.

In this review, I will argue that the scepticism behind both these statements is justified. We recently revisited the results on sedimentation in NGC 6397 stars presented by {\citet{Gratton_etal_2001}} using higher-quality VLT/FLAMES-UVES data \citep{Korn_etal_2006,Korn_etal_2007}. Element-specific abundance trends were identified which agree with atomic-diffusion predictions, {\em if\/} turbulent mixing below the convective envelope is accounted for in a parametrized way. Have we thus detected signatures of Atomic Diffusion in Old Stars (ADiOS)? Or are these trends mere artefacts of Conspiring Inaccuracies in Abundance Observations (CIAO)?

}
\maketitle{}

\section{On theory}
Since the pioneering work of \cite{Aller_Chapman_1960}, microscopic diffusion has been studied in a variety of astrophysical applications, ranging from ordinary low-mass stars like the Sun to chemically-peculiar (CP) stars, horizontal-branch stars and white dwarfs. In the latter objects, effect of atomic diffusion (that is, sedi\-mentation or levitation due to the net force atoms and ions will feel in stars) are large and well-studied (see Georges Michaud, these proceedings, and references therein). In solar-type stars, diffusive effects are considerably suppressed due to the presence of convection, but diffusive drainage of heavy elements from the convection zone can still occur.

Early models (only treating {\sl gravitational settling}, see Fig.~1) predicted sedimentation by a factor of 10 or more, in particular for metal-poor stars with shallow convection zones \citep{Michaud_etal_1984}. Apart from the potential impact on studies of Galactic chemical evolution, this prediction was relevant for the correct interpretation of the lithium abundances in warm halo stars \citep{Spite_Spite_1982} which was identified with the amount of lithium produced in Big-Bang nucleosynthesis. Observational constraints on atomic diffusion in solar-type stars could only be obtained in the mid-1990s \citep{Ryan_etal_1996} and pointed towards near-negligible corrections for lithium.

\begin{figure*}[t!]
\resizebox{0.71\hsize}{!}{\includegraphics[clip=true]{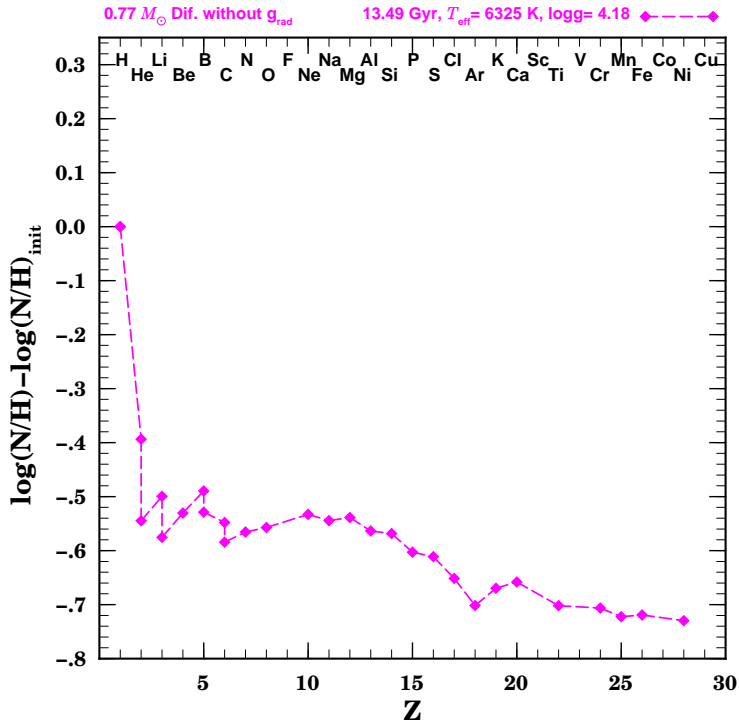}}\centering
\caption{\footnotesize
Predictions of stellar-structure models including the effect of gravitational settling for a metal-poor ([Fe/H]$_{\rm init}$ = $-2$) low-mass star that evolves to the turnoff in the course of 13.5\,Gyr. A general depletion of the atmospheric abundances of heavy elements by factors of between 3 and 5 is predicted.
}
\end{figure*}

Over the years, the theoretical predictions were refined by including additional effects. Depending on the atomic structure, the downward diffusion of elements is more or less counteracted by the interaction of the atoms with the radiation field. This process, {\sl radiative acceleration}, can levitate certain elements into the atmosphere, thereby raising their abundances above the original ones (see Fig.~2). Note that lithium is not levitated. This refinement was therefore incapable of addressing the observational fact that the {\sl Spite plateau} of lithium is observed to be thin and flat, without a downturn towards hotter and more massive stars with shallower convection zones and larger predicted diffusion corrections. It was claimed that one simply had not observed a sufficient number of stars to detect this predicted downturn \citep{Salaris_Weiss_2001}. This idea did not really catch on with the observers, in spite of the accumulating evidence for a high baryon density $\Omega_{\rm b}$ (and with it a high primordial lithium abundance) as inferred from measurements of CMB anisotropies \citep{Netterfield_etal_2002}.

\begin{figure*}[t!]
\resizebox{0.71\hsize}{!}{\includegraphics[clip=true]{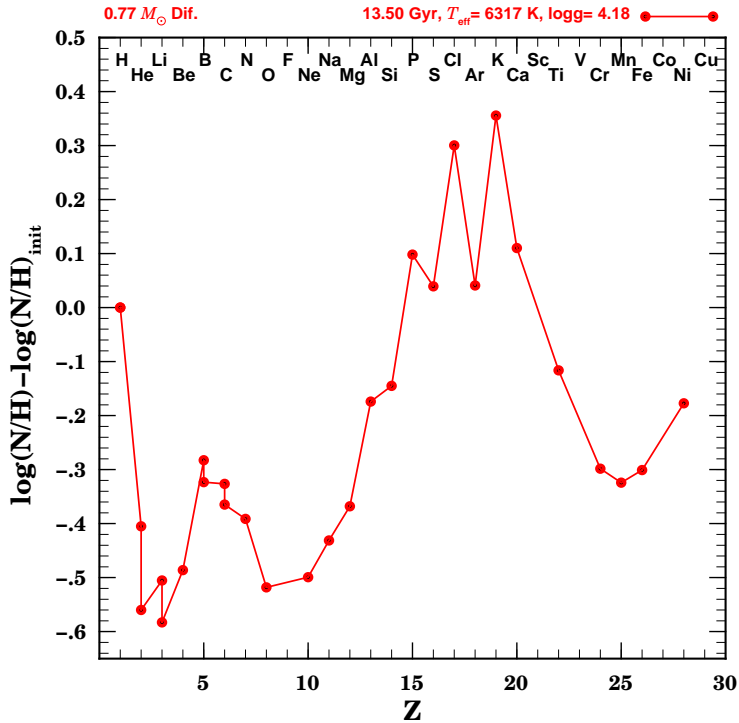}}\centering
\caption{\footnotesize
Predictions of stellar-structure models including the effects of gravitational settling and radiative acceleration for the metal-poor star of Fig.~1. Depletions of up to a factor of 4 and enhancements of up to a factor of 2.5 are predicted by these models. Note the slightly shifted ordinate scale.
}
\end{figure*}

Around the same time, it became more and more evident that additional mixing processes are required to describe the properties of Am/Fm stars. In these, radiative accelerations accumulate iron at depths with temperatures around 200\,000\,K leading to the creation of an iron convection zone \citep{Richer_etal_2000,Richard_etal_2001}. To moderate the large abundance variations produced by uninhibited atomic diffusion, one needs to introduce a mixing mechanism that keeps the layers between the iron and the helium convection zone mixed. This process is referred to as {\sl turbulent mixing}. It is parametrized by
\begin{displaymath}
D_T = 400 \,D_{\rm He}(T_0)(\frac{\rho}{\rho(T_0)})^{-3}
\end{displaymath}
where $D_T$ is the diffusion coefficient for turbulent mixing, $D_{\rm He}(T_0)$ is the isotropic atomic diffusion coefficient of helium at the reference temperature $T_0$ and $\rho$ is the density. Thus, turbulent mixing is localized to a thin layer below the convection zone. This formulation was already applied to study the effects of turbulent mixing on gravitational settling in the Sun \citep{Proffitt_Michaud_1991}. For more insight into the physics of turbulent mixing, see Suzanne Talon (these proceedings, and references therein).

\cite{Richard_etal_2005} showed that the inclusion of turbulent mixing allows to meet the observational constraint of a flat and thin {\sl Spite plateau} in the presence of significant surface depletions (see their Fig.~7): depending on the assumed strength (varying the reference temperature between log\,$T$ = 6.0 and 6.28), depletions of 0.2 to 0.6\,dex are predicted.

In Fig.~3, such predictions are presented. As indicated above, turbulent mixing moderates both the large depletions and the enhancements. Some elements are now predicted to be depleted by $\approx 0.2$\,dex (Mg, Al, Si), while others suffer a decrease of merely $\approx 0.1$\,dex (Ca, Sc, Ti). Iron-group elements fall in between these two (very moderate!) extremes. Identifying such abundance signatures is the challenge one is faced with.

\begin{figure*}[t!]
\resizebox{0.721\hsize}{!}{\includegraphics[clip=true]{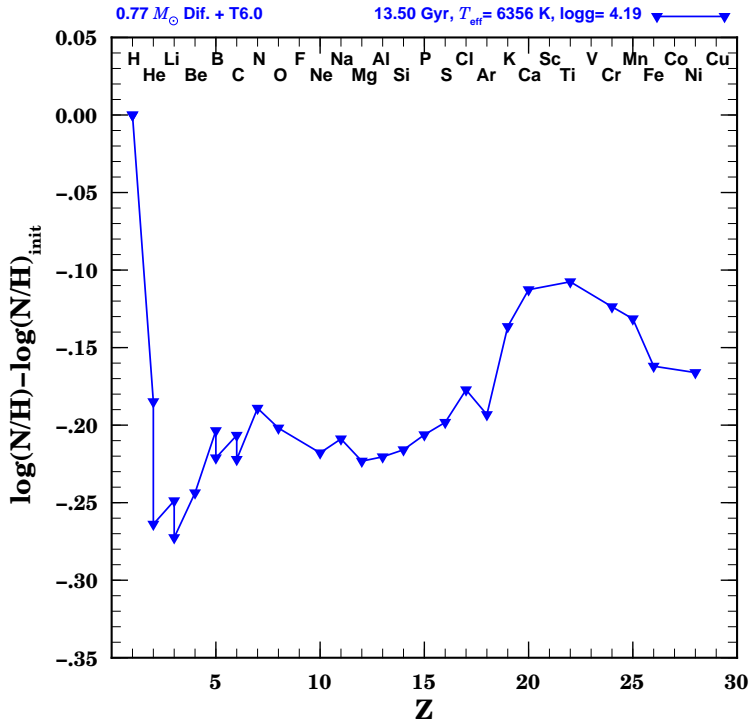}}\centering
\caption{\footnotesize
Predictions of stellar-structure models including the effects of gravitational settling, radiative acceleration and T6.0 turbulent mixing for the metal-poor star of Figs.~1 and 2. Depletions by factors of 1.25 to 2 are predicted by these models. Note the modified ordinate scale with respect to Fig.~1 and 2.
}
\end{figure*}

\section{On observations}
It was already mentioned that the observations of the {\sl Spite plateau} of lithium put some constraints on uninhibited atomic diffusion. But atomic diffusion affects all elements to some degree, while the star evolves along the main sequence (MS). When the star finally becomes a red giant and the convection zone expands inward, the original composition is practically restored. Nature supplies laboratories where the effects of atomic diffusion can be studied: globular clusters (GCs). One has to compare atmospheric abundances of stars near the MS turnoff point (TOP) to those on the red-giant branch (RGB). Were systematic abundance differences in line with the theoretical predictions (like those of Fig.~3) identified?

\begin{figure*}[t!]
\resizebox{0.98\hsize}{!}{\includegraphics[clip=true]{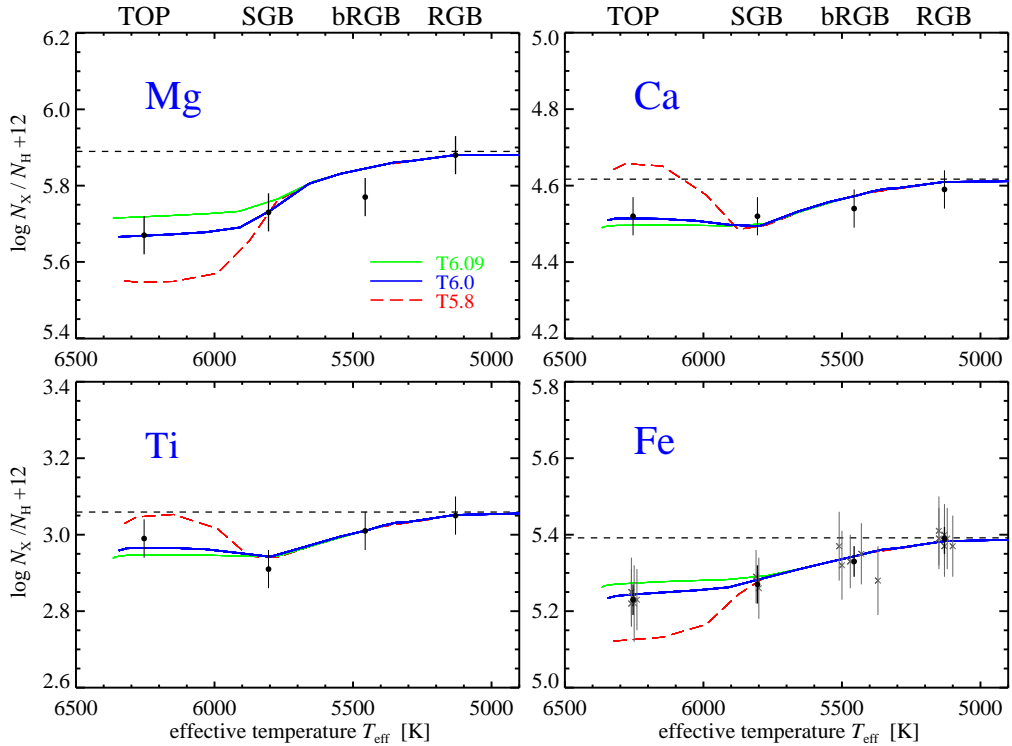}}\centering
\caption{\footnotesize
Chemical abundances (bullets and crosses) of FLAMES-UVES targets as a function of $T_{\rm eff}$. They are compared to predictions from models treating gravitational settling, radiative levitation and turbulent mixing, the latter with $T_0$ varying between log\,$T$ = 5.8 and 6.09. The intermediate T6.0 model described the element-specific trends remarkably well.
}
\end{figure*}

\cite{King_etal_1998} used the Keck telescope to observe six subgiants in M 92 ([Fe/H]\,=\,$-2.3$). While the resolving power was adequate ($R$\,=\,45\,000), the signal-to-noise ratio (S/N) was only 25-40. Because of this (and the lack of giant stars observed in the same program), they could not draw firm conclusions about abundance differences between TOP and RGB stars. They state: ``We note possible evidence for [Fe/H] differences within M92."

\cite{Gratton_etal_2001} were the first to use UVES on the VLT to study unevolved stars in metal-poor globular clusters in greater detail. The main aim of their study was to verify or refute the existence of elemental anticorrelations that had previously only been studied in GC giants. With S/N ratios around 80 at $R$\,=\,40\,000, the nominal data quality for stars in NGC 6397 ([Fe/H]\,=\,$-2$) is good, but the spectra suffer from flat-fielding and order-merging artefacts which lead to systematically overestimated TOP-star effective temperatures \citep{Korn_etal_2004,Korn_etal_2007}. As TOP and base-RGB stars were observed, the issue of atomic diffusion could also be addressed: ``[Fe/H]'s obtained for TO-stars agree perfectly with that obtained for stars at the base of the RGB."

\cite{Cohen_Melendez_2005} studied 22 RGB and three subgiant stars in M 13. They write in the abstract that ``Most elements, including Fe, show no trend with $T_{\rm eff}$ [...] suggesting that [...] gravitationally induced heavy-element diffusion [is] not important [...]". However, a closer look at the published iron abundances reveals a systematic difference between the two groups of stars:
[Fe/H]$_{\rm SGB}$\,=\,[Fe/H]$_{\rm RGB} - 0.15$.

In \cite{Korn_etal_2006,Korn_etal_2007} we presented the analysis of in total 18 stars covering the post-MS evolutionary sequence of NGC 6397 in some detail: five TOP stars, two stars in the middle of the subgiant branch, five base-RGB stars and six RGB stars. They were observed with FLAMES-UVES on the VLT with nomi\-nal data quality very similar to that achieved by \cite{Gratton_etal_2001}: $R$\,=\,48\,000, 70 $\leq$ S/N $\leq$ 110. But since FLAMES-UVES is a fibre-fed spectrograph, the flat-fielding is more reliable and the intrinsic profiles of the strong TOP-star H$\alpha$ lines can be recovered with confidence. In addition, we observed 130 stars along the subgiant branch with FLAMES-MEDUSA at somewhat lower resolution ($R$\,$\approx$\,26\,000). These spectra have a free spectral range of 200\,\AA\ which facilitates Balmer-profile analyses substantially.

Note that it suffices to derive reliable stellar-parameter {\em differences} to study abundance differences between stars. For GC stars, log $g$ differences can be accurately determined from $V$-magnitude differences (with only weak dependencies on other parameters). Here, spectroscopic techniques are less precise. For $T_{\rm eff}$ differences, we rely on both photometric and spectroscopic means. It is this combination of independent techniques that gives us confidence in our results.

\begin{table*}\centering
\caption{Stellar-parameter differences between TOP and RGB stars}
\begin{tabular}{lrcc}
\hline\\
method & $\Delta$ $T_{\rm eff}$ (TOP $-$ RGB) [K] & $\Delta$ log $g$ (TOP $-$ RGB) & $\Delta$ [Fe/H] \\
\hline
\\
H$\alpha$ & 1124 $\pm$ 120 & 1.38 $\pm$ 0.12 & 0.16 $\pm$ 0.05 \\
$V$ & & 1.38 $\pm$ 0.05 & \\
$V-I$ & 1070 $\pm$ 50 & & \\
$B-V$ & 906 $\pm$ 50 & & \\
$v-y$ & 1082 $\pm$ 50 & & \\
$b-y$ & 1108 $\pm$ 50 & & \\
\hline
\end{tabular}
\end{table*}

Based on both spectroscopic and photometric stellar parameters (see Table 1), we find trends of atmospheric abundances with evolutionary phase that are in good agreement with the predictions from models treating gravitational settling, radiative levitation and turbulent mixing (see Fig.~4). The steepest trends are identified for Mg and Fe both of which are significant at the 3$\sigma$ level. Similar trends are identified among the 100+ FLAMES-MEDUSA stars using independent techniques (Lind et al., in prep.).

\section{On primordial lithium}

\begin{figure*}[t!]
\resizebox{0.71\hsize}{!}{\includegraphics[clip=true]{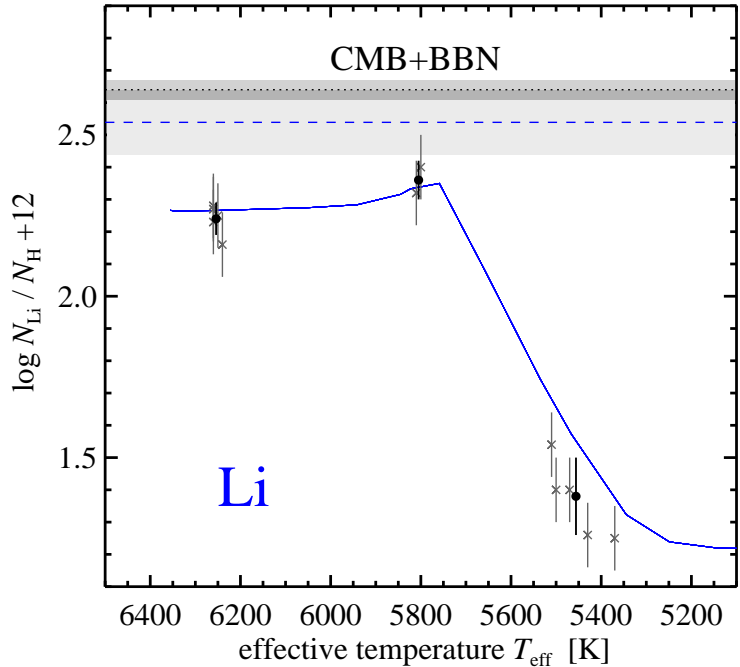}}\centering
\caption{\footnotesize
Observed lithium abundances in the TOP and subgiant stars compared to trends predicted by the T6.0 model including the effects of gravitational settling, radiative acceleration and turbulent mixing. Note that there is a diffusion signature between the TOP and subgiant stars. The light-shaded area indicates the original lithium abundances of these stars, while the dark-shaded one represents the primordial lithium abundance as inferred by WMAP \citep{Spergel_etal_2007}.
}
\end{figure*}

The observed lithium abundances of the TOP stars scatter around $\log \varepsilon$\,(Li)\,=\,2.24\,$\pm$\,0.05 (1$\sigma$). The two subgiant stars show a mean lithium abundance of $\log \varepsilon$\,(Li)\,=\,2.36\,$\pm$\,0.06. Both the base-RGB and the RGB stars show lower lithium abundances indicative of dilution with lithium-free layers, as the convection zone expands inward (see Fig.~5).

We can use the empirically calibrated atomic-diffusion model to predict the lithium abundances the NGC 6397 stars had at the time of formation. This value ($\log \varepsilon$\,(Li)$_{\rm p}$\,=\,2.54\,$\pm$\,0.1) is in good agreement with the predictions of Big-Bang nucleosynthesis at $\Omega_{\rm b}$ as derived by WMAP \citep{Spergel_etal_2007}. In \cite{Korn_etal_2006}, we called this ``A probable stellar solution to the cosmological lithium discrepancy''.

\section{On antagonism}

\cite{Bonifacio_etal_2007} criticized our work in that it ``relies heavily on the adopted temperature scale, which is plausible, albeit inconsistent with the cluster photometry. An increase by only 100\,K of the effective temperature assigned to the TO stars would remove the abundance differences [...]''. As Table 1 shows, our spectroscopic effective-temperature scale is not at all inconsistent with {\sl the cluster photometry}. Rather, it agrees within 1$\sigma$ error bars with three of the four colour indices studied and is conservative in the sense that it points to a larger $T_{\rm eff}$ difference yielding smaller abundance differences. Moreover, raising the TOP-star effective temperatures by 100\,K diminishes the abundance trends by typically 0.05 (Mg) to 0.07\,dex (Li). While this may make the small abundance difference between lithium in TOP and subgiant stars (0.12\,dex) insignificant, it only removes one quarter of the identified abundance trend for Mg.

Criticism of more general nature was raised by \cite{Asplund_etal_2006}. The detection of $^6$Li in the spectra of several metal-poor stars puts constraints on atomic diffusion and mixing, as $^6$Li is even more fragile than $^7$Li. Some of the atomic-diffusion models (those with the most efficient turbulent mixing compatible with the {\sl Spite plateau}: T6.25 and T6.28) destroy $^6$Li very efficiently lowering its abundance by up to factors of 100. Consequently, the original $^6$Li abundance (observed to be $\approx$ 5\,\% that of $^7$Li) would be {\em higher} than the $^7$Li abundance. Our observations point to a low-efficiency model (T6.0) that depletes roughly equal amounts of both isotopes (see Fig.~3).
Together with pre-MS burning, this may aggravate the $^6$Li problem. Note, however, that the reality of the $^6$Li detections are now questioned \citep{Cayrel_2007}.

\section{On diffusion in M 67}

Recently, \cite{VandenBerg_etal_2007} studied the solar-metallicity, old (4\,Gyr) open cluster M~67 using up-to-date hydrostatic model atmospheres as outer boundary conditions to the stellar-structure models. They found that the observed {\sl blueward hook} at the TOP (indicative of a transition in mass ($M_{\rm tr}$) from stars without a convective core to those with a convective core) is a sensitive function of the assumed composition and can only be reproduced by models with a high solar metallicity similar to that given by \cite{Grevesse_Sauval_1998}. Atomic diffusion may be able to lower $M_{\rm tr}$ sufficiently (by $\approx 0.04\,$M$_\odot$) to resolve this newly identified discrepancy.

\section{On the future}

We are only beginning to study the effects of atomic diffusion and mixing on atmospheric abundances of Pop II stars.
More theoretical and observational work is required to identify the physical process(es) responsible for moderating
the effects of uninhibited atomic diffusion. The inclusion of turbulent mixing in a layer just below the convective envelope
is remarkably successful in describing the observed abundance trends. The description applied is, however, in no way unique and physically
satisfying. Mass loss may be capable of producing a similar moderation (see \cite{Vauclair_Charbonnel_1995} and Mathieu Vick, these proceedings). In the coming years, the ADiOS\footnote{The ADiOS team currently consists of Martin Asplund, Paul Barklem, Lionel Bigot, Corinne Charbonnel, Remo Collet, Frank Grundahl, Bengt Gustafsson, Karin Lind, Lyudmila Mashonkina, Georges Michaud, Nikolai Piskunov, Olivier Richard, Suzanne Talon and Fr\'{e}d\'{e}ric Th\'{e}venin. The effort is led by the author. Seven of us attended this conference.}
team hopes to put additional constraints on the dependence of turbulent mixing (and mass loss) on stellar parameters, both observationally (by studying other globular clusters) and theore\-tically (by means of sophisticated (hydrodynamic) stellar-structure models). Encouraging results point to the importance of internal gravity waves \citep{Charbonnel_Talon_2005}.

Given our present ignorance about the extent of atomic diffusion as a function of, e.g., metallicity, one should exercise caution when
interpreting the abundance pattern of extreme Pop II stars like HE 1327$-$2326 \citep{Frebel_etal_2005}. While among low-mass stars atomic diffusion produces largest effects at low metallicities, it is an omnipresent phenomenon which we should strive to include in XXI.~Century modelling.

\begin{acknowledgements}
I wish to thank all the scientific friends and colleagues who have contributed to the ongoing ADiOS effort. In particular, I would like to thanks Olivier Richard who supplied some of the figures for this article.
\end{acknowledgements}

\bibliographystyle{aa}

\end{document}